# Replay spoofing detection system for automatic speaker verification using multi-task learning of noise classes

**Hye-Jin Shim**
School of Computer Science
University of Seoul
Seoul, South Korea
shimhz6.6@gmail.com

**Jee-Weon Jung**
School of Computer Science
University of Seoul
Seoul, South Korea
jeewon.leo.jung@gmail.com

**Hee-Soo Heo**
School of Computer Science
University of Seoul
Seoul, South Korea
zhasgone@naver.com

**Sung-Hyun Yoon**
School of Computer Science
University of Seoul
Seoul, South Korea
ysh901108@naver.com

**Ha-Jin Yu**
School of Computer Science
University of Seoul
Seoul, South Korea
hjyu@uos.ac.kr

*Abstract*— In this paper, we propose a replay attack spoofing detection system for automatic speaker verification using multi-task learning of noise classes. We define the noise that is caused by the replay attack as replay noise. We explore the effectiveness of training a deep neural network simultaneously for replay attack spoofing detection and replay noise classification. The multi-task learning includes classifying the noise of playback devices, recording environments, and recording devices as well as the spoofing detection. Each of the three types of the noise classes also includes a genuine class. The experiment results on the version 1.0 of ASVspoof2017 datasets demonstrate that the performance of our proposed system is improved by 30% relatively on the evaluation set.

## I. INTRODUCTION

As speaker verification is applied to various applications, the reliability of automatic speaker verification systems became an important issue. Therefore, many researchers are focusing on spoofing detection to enhance the reliability of speaker verification system. An audio spoofing signal is generated by manipulating a genuine signal through recording, synthesizing or modifying to trick a speaker verification system.

From amongst other studies in this field, automatic speaker verification(ASV) spoofing and countermeasures challenge has initiatively led to the evaluation of audio spoofing using various attacks such as speech synthesis, voice conversion in 2015 [1], and replay attack in 2017 [2], respectively. The results of the ASVspoof2017 challenge showed that replay attack is more difficult to detect than other attacks. While speech synthesis and voice conversion are hard to implement as they need special equipment and expertise, replay attack requires neither special equipment nor expertise. Additionally, speech synthesis and voice conversion also include a playback phase that occurs after manipulation of the genuine signal. Hence, detecting replay attack can help in detecting other spoofing attacks. In this paper, the spoofed signal and spoofing detection are discussed only in the context of replay attack and replay attack spoofing detection.

Generally, channel noise in an audio signal is caused by the recording environment and recording or playing devices. In various domain such as speaker recognition, it has been known that channel noise reduces the accuracy of the system [3]. Therefore, to improve the performance, many studies focus on reducing channel noise.

However, in spoofing detection, we hypothesized that noise can be vital especially in replay attack. In replay attack, the noises of the recording environment, playback and recording devices are generated during the playback and re-recording phase. Compared with the original signal without replay attack, the spoofed signal is identical to the genuine signal except the replay noise which is added by recording environment, recording and playback devices. We define this additional channel noise that is added during replay attack as replay noise.

The conventional method of spoofing detection is binary classification. This technique is used to determine whether a signal has been spoofed. Given the importance of replay noise, we trained a deep neural network (DNN) for replay noise classification as well as spoofing detection. Multi-task learning [4] is implemented to train the network on various tasks at the same time.

## II. RELATED WORKS

In the ASVspoof2017 in which the systems for detecting replay attacks were evaluated, many teams showed competitive results [5, 6, 7, 8]. Among these, the system developed by Lavrentyeva et al. (2017) [5] had the best performance. To verify effectiveness of our proposed system, we designed the architecture of our system in a manner similar to the aforementioned system [5], and modified it to apply replay noise classification. The system is composed of front-end DNN for extracting features and back-end single Gaussian model for scoring.

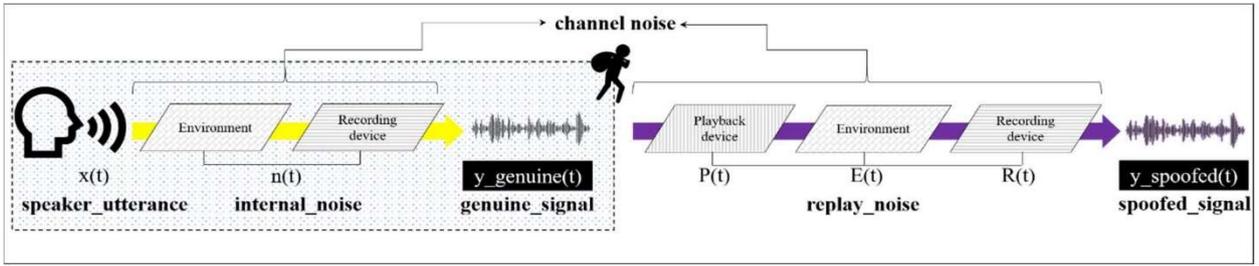

Fig. 1. Process of spoofing a signal based on the 'stolen voice' scenario

For detecting replay attack, Nagarsheth et al. (2017) [6] used a channel discrimination. The system developed by Nagarsheth et al. (2017) [6] completely trained DNN for spoofing detection or channel discrimination and selected the DNN only for channel discrimination. In our study, we trained the DNN simultaneously for spoofing detection and replay noise classification by using multi-task learning. For generalization, we added the genuine node for each replay noise classification, whereas Nagarsheth et al. (2017) did not consider the genuine class in the channel discrimination [6].

### A. Front-end DNN

The front-end DNN is used as a feature extractor. The spectrogram of a signal is fed to the network for training the spoofing detector. Spoofing detection features are extracted from the linear activation of the last hidden layer of the trained network and scoring is performed by the back-end single Gaussian model.

Given that noise is well visualized in spectrograms, a convolutional neural network (CNN) is exploited. CNNs are often used for processing images; however, recently, CNNs have performed well in spoofing detection [5, 9] as well. Among many CNN architectures, we implemented Light CNN(LCNN) [10], which showed the best performance in ASVspoof2017.

In LCNN, the concept of maxout activation [11] is applied to each CNN layer, called a Max-Feature- Map (MFM). An LCNN can select the most representative feature from various features produced by different filters. It also helps reduce the number of parameters and makes CNN faster. LCNN also shows competitive results.

### B. Back-end single Gaussian scoring

The output layer of DNN for binary classification tasks typically configured in one of the two configurations: single node with sigmoid activation, or two nodes with softmax activation. In the former configuration, the value of the node is directly used as the score and in the latter configuration, the value of a single node is used as the score.

However, the output layer's value cannot be directly used as a measure of reliability [12]. Nagarsheth et al. (2017) [5] applied single Gaussian modeling using the last hidden layer's linear activation as the code to avoid this issue.

After the DNN was trained, two single Gaussian models were modeled by calculating the mean and standard deviation of both the genuine and the spoofed signal's code, respectively. During the test phase, the code was extracted using the DNN. Then, the difference between the log probability of the genuine model and the log probability of the spoofed model was obtained and used as the score.

### III. THE PROPOSED SYSTEM

To detect a replay attack, we expect that the investigation of the difference between the genuine signal and the spoofed signal can contribute to improving the performance of the spoofing detection system. In this section, we analyze the difference between the genuine signal and the spoofed signal based on the process of producing each signal. Based on the result, we introduce a method to improve the performance of spoofing detection system.

The process of producing a genuine signal and spoofed signal based on the 'stolen voice' scenario [2] is depicted in Fig. 1. First, the genuine signal is generated when the speaker utterance is entered into the verification system through a recording device. The speaker utterance only refers to the utterance spoken by a speaker without any noise. However, internal noise is inherent in every recorded signal. In this process, the noise of the recording device and the recording environment is inevitably included in the genuine signal. Hence, we define this noise as internal noise. Next, the spoofed signal based on the 'stolen voice' scenario is produced by modifying the genuine signal which was stolen by the imposter. The imposter plays back the genuine signal and re-records it to generate the spoofed signal. The spoofed ⇄ signal includes the noise from the playback device, recording environment and the recording device as well as the genuine signal. We defined the noise of the playback device, recording environment and the recording device as replay noise. By analyzing the process of generating the genuine signal and the spoofed signal, we found that both signals commonly include the speaker utterance and the internal noise; however, only the spoofed signal includes the replay noise.

The analyzed results can be expressed by the following equations:

$$y_{genuine}(t) = x(t) * n(t) \qquad (1)$$
$$y_{spoofed}(t) = y_{genuine}(t) * P(t) * E(t) * R(t) \qquad (2)$$

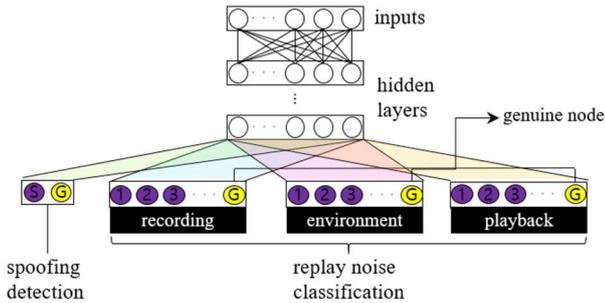

Fig. 2. System architecture of the proposed multi-task DNN

The above equations are from Alegre et al. (2014) [13]. The * operation represents the convolution. This equation represents the genuine signal and the spoofed signal as a convolution of impulse responses of various signal.

The genuine signal (1) is composed of the speaker utterance x(t) and internal noise n(t). We assumed that internal noise is not informative especially in a replay attack, as the internal noise is inevitably included in recorded signal. However, the spoofed signal (2) is composed of not only the genuine signal but also the noise from playback devices P(t), the recording environment E(t) and the recording devices R(t). The replay noise is added after spoofing and the replay noise can be the biggest difference between the spoofed signal and the genuine signal. As a result, we proposed the method of training a DNN for replay attack spoofing detection and replay noise classification at the same time.

As we intended to concurrently embed the feature of spoofing detection and replay noise classification, we applied multi-task learning. Multi-task learning is a method of learning several tasks at the same time. From learning each task, other tasks also can be learned better and a synergy effect exists between different tasks [4]. Thus, the total classes are composed of the results of spoofing detection(genuine/spoofed) and three kinds of replay noise classification. Proposed system architecture is depicted in Fig. 2. For each replay noise classification, we added a single node which indicates genuine if the signal is genuine. For instance, in the task that identifies a playback device, there are as many as nodes as the number of playback devices and one extra node indicating that the input signal is the genuine signal. We expect that by adding the genuine node for the genuine signal, the generalization ability can be improved.

TABLE I. DATA CONFIGURATION

| Subset | The number of utterances | |
|---|---|---|
| | *Non-replay* | *Replay* |
| Train | 1508 + 380 | 1508 + 570 |
| Dev | 380 | 380 |
| Eval | 1298 | 12008 |
| Total | 3566 | 14466 |

## IV. EXPERIMENTS

### A. Datasets

Experiments were conducted on the version 1.0 of ASVspoof2017 datasets, composed of non-replayed utterances (genuine signals) and replayed utterances (spoofed signals). Non-replayed utterances are subsets of original RedDots corpus and replayed utterances are made by playing back and recording the RedDots corpus for simulating a replay attack spoofing environment. The ASVspoof2017 corpus is originally divided into training, development and evaluation, but re-partitioning of training and development subsets are permitted.

Given that training a network using only training set does not involve sufficient channel information, we used part of the development set for training a DNN. The separation of the dataset is as listed in TABLE Ⅰ. We used the non-replayed utterances of the development set which are composed of 380 utterances from the original development set (760 utterances).

The 380 utterances are selected from 50 utterances of seven speakers (M11, M12, M13, M14, M16, M17) and 30 utterances of one speaker (M15). In all experiments, the inputs are spectrograms which are obtained via Fast Fourier Transform(FFT) using the kaldi [14] default setup. If the signal is longer than four seconds, we randomly select 4 seconds of the signal and if the signal is shorter than four seconds, we duplicate it. We also applied mean vector normalization at the utterance level.

### B. Baseline

To verify effectiveness of classifying the replay noise, we implemented the best performance system in [5] from the ASVspoof2017 challenge. However, the baseline was manipulated at several points and the obtained results are also different.

Difference of our baseline from the system in [5] are as follows. First, the number of dimensions and the method of extracting the spectrogram is different from those in [5]. Specifically, the size of input in [5] is (864×400×32), but the size of input in this work is (400×257×32). Second, we did not use any additional datasets, though the authors in [5] used additional datasets. Third, the separation of the development set and the training set is not same as the separation in [5]. Specific details on this were mentioned in section 5.1. Fourth, considering the different size of input spectrogram, we exploited (2×1) max-pooling operation instead of (2×2). We also added Dropout of 20% [15] after the input. Finally, ADAM optimizer [16] was used with learning rate of $10^{-3}$.

## C. The proposed system

TABLE II. THE OVERALL DNN ARCHITECTURE

| Type | Filter/Stride | Output |
|---|---|---|
| Input | | |
| Dropout1(0.2) | | |
| Conv1 | $5 \times 5 / 1 \times 1$ | $400 \times 257 \times 32$ |
| MFM1 (1/2) | - | $400 \times 257 \times 16$ |
| MaxPool1 | $2 \times 2 / 2 \times 2$ | $200 \times 128 \times 16$ |
| Conv2a | $1 \times 1 / 1 \times 1$ | $200 \times 128 \times 32$ |
| MFM2a (1/2) | - | $200 \times 128 \times 16$ |
| Conv2b | $3 \times 3 / 1 \times 1$ | $200 \times 128 \times 48$ |
| MFM2b (1/2) | - | $200 \times 128 \times 24$ |
| MaxPool2 | $2 \times 2 / 2 \times 2$ | $100 \times 64 \times 24$ |
| Conv3a | $1 \times 1 / 1 \times 1$ | $100 \times 64 \times 48$ |
| MFM3a (2/3) | - | $100 \times 64 \times 32$ |
| Conv3b | $3 \times 3 / 1 \times 1$ | $100 \times 64 \times 64$ |
| MFM3b (1/2) | - | $100 \times 64 \times 32$ |
| MaxPool3 | $2 \times 1 / 2 \times 1$ | $50 \times 64 \times 32$ |
| Conv4a | $1 \times 1 / 1 \times 1$ | $50 \times 64 \times 64$ |
| MFM4a (1/2) | - | $50 \times 64 \times 32$ |
| Conv4b | $3 \times 3 / 1 \times 1$ | $50 \times 64 \times 32$ |
| MFM4b (1/2) | - | $50 \times 64 \times 16$ |
| MaxPool4 | $2 \times 1 / 2 \times 1$ | $25 \times 64 \times 16$ |
| Conv5a | $1 \times 1 / 1 \times 1$ | $25 \times 64 \times 32$ |
| MFM5a (1/2) | - | $25 \times 64 \times 16$ |
| Conv5b | $3 \times 3 / 1 \times 1$ | $25 \times 64 \times 32$ |
| MFM5b (1/2) | - | $25 \times 64 \times 16$ |
| MaxPool5 | $2 \times 2 / 2 \times 2$ | $13 \times 32 \times 16$ |
| Dropout2(0.7) | | |
| FC6 | | $2 \times 64$ |
| FC7 | | $2 \times 64$ |
| FC_S | | 2 |
| FC_E | | 4+1 |
| FC_P | | 8+1 |
| FC_R | | 7+1 |

TABLE III. DETAILS OF MULTI-TASK NODES

| Object of classifier | Task details | # of classes |
|---|---|---|
| Spoofing detection (FC_S) | Spoofed and genuine | 2 |
| Recording environment (FC_E) | ['Balcony', 'Bedroom', 'Cantine', 'Office'] (and genuine) | 4+1 |
| Playback device (FC_P) | ['All-in-one PC speakers', 'Beyerdynamic DT 770 PRO headphones', 'Creative A60', 'Dell laptop with internal speakers', 'Dynaudio BM5A Speaker connected to laptop', 'HP Laptop speakers', 'High Quality GENELEC Studio Monitors Speakers', 'VIFA M10MD-39-08 Speaker connected to laptop'] (and genuine) | 8+1 |
| Recording device (FC_R) | ['BQ Aquaris M5 smartphone. Software: Smart voice recorder', 'Desktop Computer with headset and arecord', 'H6 Handy Recorder', 'Nokia Lumia', 'Rode NT2 microphone connected to laptop', 'Rode smartlav+ microphone connected to laptop', 'Samsung Galaxy 7s'] (and genuine) | 7+1 |

TABLE IV. PERFORMANCE COMPARISON OF SPOOFING DETECTION EER(%) WITH THE BEST PERFORMANCE SYSTEM

| System | Zhang, C. et al | | Baseline | | Proposed | |
|---|---|---|---|---|---|---|
| Set | Valid | Eval | Valid | Eval | Valid | Eval |
| EER (%) | - | 11.50 | 9.47 | 13.57 | 4.21 | 9.56 |

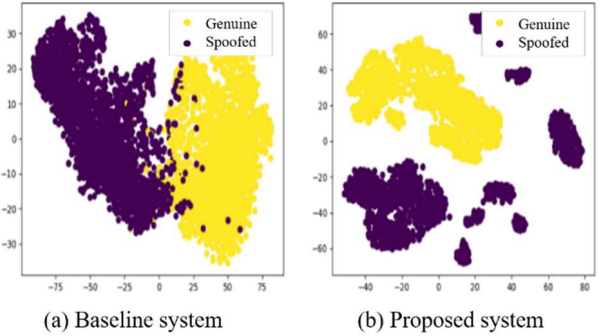

(a) Baseline system    (b) Proposed system

Fig. 3. Visualization of t-SNE results compared to baseline system

We implemented the multi-task learning for concurrently training spoofing detection and the replay noise classification. We trained the network with four tasks as follows and the output layers were eliminated after training. It must be noticed that the output layers are composed of not only the results of spoofing detection (genuine/spoofed) and the classes of all kind of replay noise, but also a genuine node for each replay noise classification. Thus, the total number of output layer nodes is 2 + (4+1) + (8+1) + (7+1) = 24. All losses in multi-task learning are equally weighted. Other DNN configuration details are listed in the TABLE II, FC and MFM refers to Fully Connected Layer and Max Feature Map, respectively. With regard to the multi-task output nodes which is listed in the last two columns of TABLE Ⅱ, the details are included in TABLE Ⅲ.

## D. Results and Disccussion

The experimental results are listed in the TABLE Ⅳ. The results show that the usage of replay noise can improve the performance of spoofing detection systems. The different performances in the valid set and eval set is because the replay noise of the valid set also exists in the training set. However, the performance improvement on the eval set proves the effectiveness of the replay noise when the devices and environment are even unknown.

Fig. 3 shows the visualization results using t-SNE [17]. In this figure, the yellow dots and purple dots indicate the genuine and the spoofed signal, respectively. From the visualization results, we found that the distribution of the spoofed signal is represented by multiple clusters which are caused by the channel difference and the overlap with the genuine signal is significantly reduced by training for the classification of the replay noises.

## V. CONCLUSION

In this paper, we proposed a reply attack spoofing detection system using multi-task learning with the replay noise classes. To verify its effectiveness, we conducted an experiment on the version 1.0 of ASVspoof2017 datasets. We trained a DNN to perform the replay noise classification task as well as spoofing detection task by multi-task learning. Replay noise classification is composed of three sub tasks which classify the noise as environments, recording devices and playback devices, respectively. Experimental results show the improvement in the spoofing detection performance. EER of the proposed system is reduced by about 30% relatively on the evaluation set.